\documentclass[letterpaper,12pt]{article}
\usepackage{osajnl2}
\usepackage{amsfonts, bbm, amsfonts}
\usepackage[draft]{hyperref}

\begin{document}

\title{Filtering equation for a measurement of a coherent channel}

\author{Anita D\c{a}browska$^{*}$ and Przemys{\l}aw Staszewski}
\address{Department of Theoretical Foundations of Biomedical Sciences
and Medical Informatics\\ Collegium Medicum, Nicolaus Copernicus
University\\  ul.~Jagiello\'nska 15, 85-067 Bydgoszcz, Poland }
\address{$^*$Corresponding author: adabro@cm.umk.pl}

\begin{abstract}
A stochastic model for a continuous photon counting and heterodyne
measurement of a coherent source is proposed. A nonlinear
filtering equation for the posterior state of a single-mode field
in a cavity is derived by using the methods of quantum stochastic
calculus. The posterior dynamics is found for the observation of a
Bose field being initially in a coherent state. The filtering
equations for counting and diffusion processes are given.
\end{abstract}

\maketitle

\section{Introduction}

The quantum filtering equation describes the dynamics of a quantum
system undergoing an indirect observation continuous in time
\cite{Bel89, Bel90, BarBel91, Bel92a, BelSta92, Bel99, BGM04,
WisMil00}. In the model based on the quantum stochastic calculus
(QSC) \cite{HudPar84, GarCol85, Par92, Gar00}, worked out by
Hudson and Parthasarathy, a role of measuring apparatus is played
by a Bose field \cite{BraRob81}. The dynamics of a compound system
consisting of the quantum system in question and the Bose field is
given by a unitary operator. The external field is here not only a
source of a noise but it also drives the system and carries the
information on it \cite{GarCol85}. The time development of the
posterior state conditioned by a trajectory of the results of
continuous measurement of the output field is given by the
stochastic and irreversible equation \cite{BGM04, BelSta92,
WisMil00, WisMil93a, WisMil93b, Car93}.

In this paper we apply the theory of measurement continuous in
time, based on QSC, to derive the quantum filtering equation for a
single-mode field in a cavity interacting with the Bose field
prepared in a coherent state. To get the posterior dynamics of the
system we use the generating functional approach described in
\cite{Bel89, Bel90}, but, in contrast to the derivation given by
Belavkin when the Bose field is initially in the vacuum state, we
consider the situation when the information about the system is
extracted from a measurement of a coherent channel. We obtain the
nonlinear filtering equation for a direct and heterodyne
measurement of the signal escaping from cavity.

The statistics of the output process for the Bose field in a
coherent state was considered by Barchielli in \cite{Bar90,
Bar06}, and for a mixture of coherent states by Barchielli and
Pero in \cite{BarPer02}, but the estimation of the state of the
indirectly observed quantum system was not given there.

\section{Quantum stochastic calculus}

In this section we present some basic ideas and notions of QSC in
the boson Fock space \cite{HudPar84, Par92}, which will be needed
in our paper.

We model the electromagnetic field in the symmetric (Bose) Fock space  $\mathcal{F}$ over the Hilbert
space $\mathcal{K}=\mathbb{C}^{n}\otimes L^{2}(\mathbb{R}_{+})$ of all
quadratically integrable functions from $\mathbb{R}_{+}$ into
$\mathbb{C}^{n}$,
\begin{equation}
\mathcal{F} \;=\; \mathbb{C} \oplus
\left(\bigoplus_{k=1}^{\infty}\mathcal{K}^{{\otimes}_{s}k}\right)\,.
\end{equation}
The space $\mathcal{K}$ called `one-particle space' relates to the
states of the photons. For every $f\in \mathcal{K}$ we define the
exponential vector $e(f) \in \mathcal{F}$ by
\begin{equation}
e(f) \;=\; \left(1,f,(2!)^{-1/2}f\otimes f,(3!)^{-1/2}f\otimes
f\otimes f,\ldots\right)\,.
\end{equation}
The inner product of exponential
vectors in $\mathcal{F}$ takes the form
\begin{equation}\label{product}
\langle e(g)|e(f)\rangle_{\mathcal{F}}\;=\;\exp\langle g|
f\rangle_{\mathcal{K}} \equiv \exp\left[\sum_{j=1}^{n}
\int_{0}^{\infty}\overline{g_{j}(t)}f_{j}(t)\,
\mathrm{d}t\right]\,.
\end{equation}
The normalized vectors
\begin{equation}
\iota(f) \;=\; \exp\left( -\frac{1}{2}
||f||_{\mathcal{K}}^{2}\right)e(f)\,,
\end{equation}
are called coherent vectors. In particular the coherent vector
$\iota(0) =(1,0,0,\ldots) \in \mathcal{F}$ is the vacuum state. The
coherent state $\iota(f)$ can represent the laser light.

The linear span $\mathcal{D}$ of all the exponential vectors is
dense in $\mathcal{F}$. On the dense domain $\mathcal{D}$ we
define the annihilation $A_{j}(t)$, creation $A_{j}^{\dagger}(t)$
and number $\mathit{\Lambda}_{ij}(t)$ processes as follows
\cite{HudPar84, Par92}:
\begin{equation}\label{processes}
A_{j}(t)e(f) \;=\; \int_{0}^{t}f_{j}(s){\mathrm d}s\ e(f)\,,
\end{equation}
\begin{equation}
A_{j}^{\dagger}(t)e(f)\;=\; \left.\frac{\partial}{\partial \epsilon_{j}}
e\big(f+\epsilon \chi_{[0,t)}\big)\right|_{\epsilon=0}\,,
\end{equation}
\begin{equation}
\mathit{\Lambda}_{ij}(t)e(f)\;=\; \left.-\mathrm{i}\frac{\mathrm{d}}
{\mathrm{d}\lambda} e\big(\exp(\mathrm{i}\lambda P_{ij}\chi_{[0,t)})f
\big) \right|_{\lambda=0}\,,
\end{equation}
where $\chi_{[0,t)}$ is the indicator function of $[0,t)$,
$\epsilon\equiv (\epsilon_{1}, \dots, \epsilon_{n})\in
\mathbb{R}^{n}$, $\lambda \in \mathbb{R}$ and
$(P_{ij}f)_{k}=\delta_{ik}f_{j}$.

\noindent The operators $A_{j}(t)$, $A_{j}^{\dagger}(t)$,
$\mathit{\Lambda}_{ij}(t)$ satisfy the commutation relations of
the form
$$[A_{i}(t),A_{j}(t^{\prime})]\;=\; [A_{i}^{\dagger}(t),A_{j}^{\dagger}
(t^{\prime})] \;=\;0\,,\;\;\;[\mathit{\Lambda}_{ij}(t),
\mathit{\Lambda}_{kl}(t^{\prime})]\;=\;\delta_{jk}\mathit{\Lambda}_{il}
(t\wedge t^{\prime})-\delta_{il}\mathit{\Lambda}_{kj}(t\wedge
t^{\prime}) \,, $$
\begin{equation}\label{canrel}
[A_{i}(t),A_{j}^{\dagger}(t^{\prime})]\;=\;\delta_{ij}\, t\wedge
t^{\prime}\,,\;\;[A_{j}(t),\mathit{\Lambda}_{kl}(t^{\prime})]\;=\;
\delta_{jk}\,A_{l} (t\wedge t^{\prime})\,,\;\;
[\mathit{\Lambda}_{kl}(t),A_{j}^{\dagger}(t^{\prime})]
\;=\;\delta_{lj}\,A_{k}^{\dagger}(t\wedge t^{\prime})\,,
\end{equation}
where $t \wedge t^{\prime}=\mathrm{min}(t,t^{\prime})$.

The Fock space $\mathcal{F}$ has a continuous
tensor product structure,~i.e.
\begin{equation}\label{factorisation}
\mathcal{F}\;=\;\mathcal{F}_{[0,t)} \otimes \mathcal{F}_{[t,\infty)}\,,
\end{equation}
where $\mathcal{F}_{[0,t)}$ and $\mathcal{F}_{[t,\infty)}$ are the
symmetric Fock spaces over $\mathbb{C}^{n}\otimes L^{2}([0,t))$
and $\mathbb{C}^{n}\otimes L^{2}([t,\infty))$, respectively. Let
$\mathcal{H}$ be a Hilbert space associated with some quantum system.
The family\break $\{M(t), \,t\geq 0\}$ of operators on $\mathcal{H}
\otimes \mathcal{F}$ is called a quantum adapted process if $M(t)$ acts
as the identity operator in $\mathcal{F}_{[t,\infty)}$ and can act in
non-trivial way in $\mathcal{H}\otimes \mathcal{F}_{[0,t)}$
\cite{HudPar84}.

Making use of the factorisation property (\ref{factorisation}),
Hudson and Parthasarathy gave a rigorous meaning to the quantum
stochastic equations (QSDE) of the type  \cite{HudPar84, Par92}
\begin{equation}\label{QSC}
\mathrm{d}M(t)\;=\;\sum_{j=1}^{n}\left(\sum_{i=1}^{n} F_{ji}(t)
\,\mathrm{d} \mathit{\Lambda}_{ji}(t)+E_{j}(t)\,
\mathrm{d}A_{j}(t)+D_{j}(t)\,
\mathrm{d}A_{j}^{\dagger}(t)\right)+C(t)\mathrm{d}t\,,
\end{equation}
where $M(t)$, $F_{ji}(t)$, $E_{j}(t)$, $D_{j}(t)$, $C(t)$ are
adapted processes on $\mathcal{H}\otimes \mathcal{F}$. The
increments ${\mathrm d}A_{j}(t)$, ${\mathrm d}A_{j}^{\dagger}(t)$,
${\mathrm d}\mathit{\Lambda}_{ij}(t)$ acting in
$\mathcal{H}\otimes\mathcal{F}_{[t,t+{\mathrm d}t)}$ commute with
any adapted process $N(t)$ in $\mathcal{H}\otimes\mathcal{F}$.

To work with these equations we use the quantum Ito's rule which
reads as follows. If $M^{\prime}(t)$ is the process which
satisfies an equation of the type (\ref{QSC}), then the
differential of the product $M(t)M^{\prime}(t)$ is given by the
formula \cite{HudPar84, Par92}
\begin{equation}\label{diff}
{\mathrm d} \big(M(t)M^{\prime}(t)\big)= {\mathrm d}M(t)M^{\prime}(t)
+M(t) {\mathrm d}M^{\prime}(t)+{\mathrm d}M(t){\mathrm d}
M^{\prime}(t)\,,
\end{equation}
where ${\mathrm d}M(t)\,{\mathrm d}M^{\prime}(t)$ can be computed
with the help of the multiplication table
$${\mathrm d}A_{i}(t)\,{\mathrm d}A_{j}^{\dagger}(t)\;=\;\delta_{ij}\,
{\mathrm d}t\,,\;\;\; {\mathrm d}A_{i}(t)\,{\mathrm d}
\mathit{\Lambda}_{kj}(t) \;=\;\delta_{ik}\,{\mathrm
d}A_{j}(t)\,,$$
\begin{equation}\label{Itotable}
{\mathrm d}\mathit{\Lambda}_{kj}(t)\,{\mathrm d}A_{i}^{\dagger}(t)
\;=\;\delta_{ji}\, {\mathrm d}A_{k}^{\dagger}(t)\,,\;\;\; {\mathrm
d}\mathit{\Lambda}_{ij}(t)\,
{\mathrm d}\mathit{\Lambda}_{kl}(t)\;=\;\delta_{jk}\,{\mathrm d}
\mathit{\Lambda}_{il}(t)\,,
\end{equation}
and all other products vanish.

\section{Filtering equation for photon counting measurement}

Let us consider a single-mode field in a cavity (system
$\mathcal{S}$) interacting with the Bose field. We assume that the
unitary evolution operator $U(t)$ of this compound system
satisfies the QSDE of the form
\begin{equation}\label{QSDE1}
\mathrm{d}U(t)\;=\;\left(\sqrt{\mu}\, b\, \mathrm{d}A^{\dagger}(t)
-\sqrt{\mu}\,  b^{\dagger}\, \mathrm{d}A(t)- {\mu \over 2}\
b^{\dagger}b\, \mathrm{d}t-{\mathrm{i}\over \hbar }
H\,\mathrm{d}t\right)U(t)\,, \;\;\; U(0)\;=\;I\,,
\end{equation}
where $b$ stands for an annihilation operator, $H=\hbar
\omega\left(b^{\dagger}b+{1\over 2}\right)$ is the hamiltonian of
$\mathcal{S}$ and $\mu \in \mathbb{R}$ is a real coupling
constant. The equation (\ref{QSDE1}) is written in the interaction
picture with respect to the free dynamics of the Bose field
\cite{Bar90, Bar06}. The field described by the
operators $A(t)$, $A^{\dagger}(t)$, and $\mathit{\Lambda}(t)$ is
called the input field \cite{GarCol85, Bar86} that is the field
before the interaction with the system $\mathcal{S}$. The output
processes, carrying the information about $\mathcal{S}$, are given by
the formulae:
\begin{equation}\label{output}
A^{\mathrm{out}}(t)\;=\;U^{\dagger}(t)A(t)U(t)\,,\;\;\;
\mathit{\Lambda}^{\mathrm{out}}(t)\;=\; U^{\dagger}(t)
\mathit{\Lambda}(t)U(t)\,.
\end{equation}

In this section we shall consider a direct detection of the output
signal
\begin{equation}\label{signal}
\mathit{\Lambda}^{\mathrm{out}}(t)\;=\;\int_{0}^{t} \left(
\mathrm{d}\mathit{\Lambda}(t^{\prime})+\sqrt{\mu}\,b_{t^{\prime}}\,
\mathrm{d}A^{\dagger}(t^{\prime})
+\sqrt{\mu}\,b_{t^{\prime}}^{\dagger}\,\mathrm{d}
A(t^{\prime})+\mu\, b_{t^{\prime}}^{\dagger}\,
b_{t^{\prime}}\,\mathrm{d}t^{\prime} \right)\,,
\end{equation}
where $b_{t}\,=U^{\dagger}(t)b\,U(t)$. The formula (\ref{signal})
can be obtained from (\ref{output}) and (\ref{QSDE1}) by applying
the rules of QSC. Note that the self-adjoint process
$\mathit{\Lambda}^{\mathrm{out}}(t)$ satisfies the
non-demolition principle \cite{Bel89, Bel90}
\begin{equation}\label{nondemol}
[\mathit{\Lambda}^{\mathrm{out}}(t^{\prime}), U^{\dagger}(t)
\,Z\,U(t)]\;=\;0\;\;\;\;\;\; 0\leq t^{\prime}\leq t\,
\end{equation}
for any observable $Z$ of the system $\mathcal{S}$, and the
commutativity condition
\begin{equation}
[\mathit{\Lambda}^{\mathrm{out}}(t), \mathit{\Lambda}^{\mathrm{out}}
(t^{\prime}) ]\;=\; 0\;\;\;\;\;\; \forall\,
t,\,t^{\prime}\geq 0\,.
\end{equation}
Therefore $\mathit{\Lambda}^{\mathrm{out}}(t)$ can be represented
by a classical random measure on $\mathbb{R}$ with values in
$\{0,1,2,\dots\}$.

Let us suppose that one realizes a measurement of the output process
(\ref{signal}) for the Bose field initially prepared in the coherent
state $\iota(f)$. To get the filtering equation corresponding to
the observation of a coherent channel, we introduce the generating
map, $\mathrm{g}(k,t)$, defined by
$$\mathrm{g}(k,t): Z\rightarrow \mathrm{g}(k,t)[Z]\,,$$
\begin{equation}\label{genmap}
\langle\psi |g(k,t)[Z]\psi\rangle\;=\;\langle\psi\otimes \iota(f)|
\mathrm{G}^{\mathrm{out}}(k,t)Z_{t}\psi\otimes \iota(f)\rangle\,,
\end{equation}
where
\begin{equation}\label{G}
\mathrm{G}^{\mathrm{out}}(k,t)\;=\;\exp\bigg\{\int\limits_{0}^{t}
\ln k(t^{\prime})\,\mathrm{d}{\mathit{\Lambda}}^{\mathrm{out}}
(t^{\prime})\bigg\}\,,
\end{equation}
$Z_{t}=U^{\dagger}(t)ZU(t)$ is the Heisenberg operator of $\mathcal{S}$,
$\psi$ stands for the initial state of $\mathcal{S}$, and $k$ is a
complex measurable test function satisfying $0<|k|\leq 1$.

An explicit expression for the generating map (\ref{genmap}) can
be obtained by solving the differential equation for
$\mathrm{g}(k,t)$. In order to find this equation, one should
first obtain the QSDE for the operator
$\pi^{\mathrm{out}}_{k}(t,Z)=U^{\dagger}(t)\mathrm{G}(k,t)ZU(t)$,
where
$\mathrm{G}(k,t)=U(t)\mathrm{G}^{\mathrm{out}}(k,t)U^{\dagger}(t)$.
By the rules of QSC, one can easily check that
\begin{equation}\label{QSDE2}
\mathrm{d}\mathrm{G}(k,t)\;=\;(k(t)-1)\,\mathrm{d}\mathit{\Lambda}
(t)\mathrm{G}(k,t)\,.
\end{equation}
From (\ref{QSDE1}) and (\ref{QSDE2}), one obtains
\begin{eqnarray}\label{QSDE3}
\lefteqn{\mathrm{d}\pi^{\mathrm{out}}_{k}(t,Z)
\:=\;-\big(K_{t}^{\dagger} \pi^{\mathrm{out}}_{k}(t,Z)+
{\pi}^{\mathrm{out}}_{k}(t,Z)K_{t}\big) \mathrm{d}t+ k(t)\,\mu
b_{t}^{\dagger}{\pi}^{\mathrm{out}}_{k}(t,Z)b_{t}
\mathrm{d}t}\nonumber \\ && +
\sqrt{\mu}\big(b_{t}^{\dagger}{\pi}^{\mathrm{out}}_{k}(t,Z)-
{\pi}^{\mathrm{out}}_{k}(t,Z)b^{\dagger}_{t}\big)\mathrm{d}A(t)+
\sqrt{\mu}\big({\pi}^{\mathrm{out}}_{k}(t,Z)b_{t} - b_{t}
{\pi}^{\mathrm{out}}_{k}(t,Z)\big)\mathrm{d}A^{\dagger}(t)
\nonumber \\&& +\big(k(t)-1\big)\big({\pi}^{\mathrm{out}}_{k}(t,Z)
\mathrm{d}\mathit{\Lambda}(t)+ \sqrt{\mu}\, b_{t}^{\dagger}
{\pi}^{\mathrm{out}}_{k}(t,Z)\mathrm{d}A(t)+
\sqrt{\mu}\,{\pi}^{\mathrm{out}}_{k}(t,Z)b_{t}
\mathrm{d}A^{\dagger}(t)\big)\,,
\end{eqnarray}
where $K_{t}=U^{\dagger}(t)KU(t)$ and $K=\frac{\mathrm{i}}
{\hbar}H+\frac{\mu}{2}b^{\dagger}b$.

By taking the mean value of both sides of the equation
(\ref{QSDE3}) with respect to the state\break $\eta=\psi\otimes
\iota(f)$, one gets the differential equation for the mean value
of $\pi^{\mathrm{out}}_{k}(t,Z)$:
\begin{eqnarray}
\lefteqn{\langle\mathrm{d}\pi_{k}^{\mathrm{out}}(t,Z)\rangle \:=\;
\langle\eta(t)|-\big( K^{\dagger}\pi_{k}(t,Z)+
{\pi}_{k}(t,Z)K\big)+ k(t)\,\mu b^{\dagger} {\pi}_{k}(t,Z)b}\nonumber \\
&& + \sqrt{\mu}\big(b^{\dagger}{\pi}_{k}(t,Z)-
{\pi}_{k}(t,Z)b^{\dagger}\big)f(t)+ \sqrt{\mu} \big({\pi}_{k}(t,Z)b -
b\,{\pi}_{k}(t,Z)\big)\overline{f(t)} \nonumber \\&&
+\big(k(t)-1\big)\big({\pi}_{k}(t,Z)|f(t)|^2+ \sqrt{\mu}\,
b^{\dagger}{\pi}_{k}(t,Z)f(t)+ \sqrt{\mu}\,{\pi}_{k}(t,Z)b
\overline{f(t)}\big)|\eta(t)\rangle\mathrm{d}t\,,
\end{eqnarray}
where $\pi_{k}(t,Z)=\mathrm{G}(k,t)Z$ and $\eta(t)=U(t)\eta$. Hence, the
generating map $\mathrm{g}(k,t)$ satisfies the differential equation
\begin{eqnarray}\label{diffmap}
\frac{\mathrm{d}}{\mathrm{d}t}\mathrm{g}(k,t)[Z] \;=\; \mathrm{g}(k,t)
\big[ -K^{\dagger}Z-ZK- \sqrt{\mu}\,\big(Zb^{\dagger}f(t)+
bZ\overline{f(t)}\big)- Z|f(t)|^{2}\big.\nonumber \\
\big.+k(t)\big(\mu\, b^{\dagger}Z\,b+\sqrt{\mu}\,b^{\dagger}Zf(t)+
\sqrt{\mu}\, Z\,b\overline{f(t)}+Z|f(t)|^{2}\big)\big]\,
\end{eqnarray}
with the initial condition $\mathrm{g}(k,0)[Z]=Z$.

The solution to eq. (\ref{diffmap}) is given by the von Neumann-Dyson
series
\begin{equation}\label{series1}
\mathrm{g}(k,t)[Z] \;=\; \sum_{n=0}^{\infty}\,\int\limits_{0}^{t}
\mathrm{d}t_{n}
\int\limits_{0}^{t_{n}}\mathrm{d}t_{n-1} \ldots\int\limits_{0}^{t_{2}}
\mathrm{d}t_{1}\, k(t_{1})\dots k(t_{n})
\times \nonumber
\end{equation}
\begin{equation}
\;\;\;\;\;\;\;\;\;\;\;\;\times S^{\dagger}(t_{1})\ldots S^{\dagger}
(t_{n})Z(t)S(t_{n})\dots S(t_{1})\,,
\end{equation}
where
\begin{equation}
Z(t)\;=\;\mathrm{e}^{-L^{\dagger}(t)} Z\, \mathrm{e}^{-L(t)}\,,
\end{equation}
\begin{equation}
S(t)\;=\;\mathrm{e}^{L(t)}\bigg( \sqrt{\mu}\, b+f(t)\bigg)
\mathrm{e}^{-L(t)}\,,
\end{equation}
with
\begin{equation}
L(t)\;= \;Kt+\int\limits_{0}^{t}\,\left(\sqrt{\mu}\,b^{\dagger}
\,f(t^{\prime})+
\frac{|f(t^{\prime})|^{2}}{2} \right)\, \mathrm{d}t^{\prime}\,.
\end{equation}
Let $\tau=(t_{1},t_{2},\ldots,t_{n})$ be the trajectory of the
observed counting process $\mathit{\Lambda}^{\mathrm{out}}(t)$ up
to $t$ and $\Sigma^{t}$ is a set of all finite chains $|\tau|=n\in
\{0,1,2,\ldots\}$. By introducing the stochastic operator\break
$V(\tau\,|\,t)\;=\;\mathrm{e}^{-L(t)}S(t_{n})\,\ldots \,S(t_{1})$,
one can rewrite the series (\ref{series1}) in the form
\begin{equation}
\mathrm{g}(k,t)[Z] \;=\; \int\limits_{\tau\in \Sigma^{t}}k(\tau)
V^{\dagger} (\tau\,|\,t)\,ZV(\tau\,|\,t)\,\mathrm{d}\tau\,,
\end{equation}
where $k(\tau)=\prod\limits_{i=1}^{n}k(t_{i})$ and  $\mathrm{d}\tau=
\prod\limits_{i=1}^{n}\mathrm{d}t_{i}$. The stochastic propagator
$\widehat{V}(t)(\tau)=V(\tau\,|\,t)$ defining for any trajectory
$\tau$ the posterior state
$\widehat{\psi}(t)(\tau)={V}(\tau|t)\psi$ of $\mathcal{S}$,
can be represented by the Ito chronological multiplicative integral
\begin{equation}\label{propa}
\widehat{V}(t)\;=\;\mathrm{e}^{-L(t)}
\sum_{n=0}^{\infty}\,\int\limits_{0}^{t}\mathrm{d}t_{n}
\int\limits_{0}^{t_{n}}\mathrm{d}t_{n-1} \ldots\int\limits_{0}^{t_{2}}
\mathrm{d}t_{1}\, S^{\prime}(t_{n})\,\ldots \,
S^{\prime}(t_{1})\prod_{i=1}^{n} \mathrm{d}N(t_{i})\,.
\end{equation}
In the last formula
\begin{equation}
S^{\prime}(t)\;=\;\mathrm{e}^{L(t)}
\bigg(\sqrt{\mu}\,b+f(t)-I\bigg)\mathrm{e}^{-L(t)}\,
\end{equation}
and $N(t)$ is a random variable such that for $\tau\in \Sigma^{\infty}$
one has $\mathrm{d}N(t)(\tau)=1$ if
$t\in \tau$ and $\mathrm{d}N(t)(\tau)=0$ if $t \notin \tau$.

By differentiation of (\ref{propa}), we get the linear stochastic
equation
\begin{equation}
\mathrm{d}\widehat{V}(t)\;=\;-\bigg(K+\sqrt{\mu}\,b^{\dagger}f(t)+
\frac{|f(t)|^2}{2} \bigg) \widehat{V}(t)\,\mathrm{d}t+
\bigg(\sqrt{\mu}\,b+f(t)-I\bigg) \widehat{V}(t)\, \mathrm{d}N(t)
\,,\;\;\;\widehat{V}(0)=I\,.
\end{equation}
Thus, we draw the conclusion that the posterior unnormalized wave
function $\widehat{\psi}(t)=\widehat{V}(t)\psi$ satisfies the
linear Belavkin equation of the form
\begin{equation}\label{fillinear1}
\mathrm{d}\widehat{\psi}(t)\;=\;-\bigg(K+\sqrt{\mu}\,b^{\dagger}\,f(t)
+\frac{|f(t)|^{2}}{2} \bigg) \widehat{\psi}(t)\,\mathrm{d}t+
\bigg(\sqrt{\mu}\,b+f(t)-I\bigg) \widehat{\psi}(t)
\,\mathrm{d}N(t)\,,\;\;\;\widehat{\psi}(0)\;=\;\psi.
\end{equation}

The nonlinear filtering equation preserving the normalization of
the posterior wave function can be calculated from eq.
(\ref{fillinear1}). To derive the differential equation for
$\widehat{\varphi}(t)=\langle\widehat{\psi}(t)
|\widehat{\psi}(t)\rangle^{-1/2}\widehat{\psi}(t)$ one has to
check that
\begin{eqnarray}\label{non1}
\mathrm{d}\left(\langle\widehat{\psi}(t)|\widehat{\psi}(t)\rangle\right)
\;&=&\; -\langle\widehat{\psi}(t)|
\bigg(\sqrt{\mu}b^{\dagger}+\overline{f(t)}\bigg) \bigg(\sqrt{\mu}b+f(t)
\bigg)|\widehat{\psi}(t)\rangle\, \mathrm{d}t\,\nonumber\\
&&+\,\langle\widehat{\psi}(t)| \left[\bigg(\sqrt{\mu}b^{\dagger}+
\overline{f(t)} \bigg)
\bigg(\sqrt{\mu}b+f(t)\bigg)-1\right]|\widehat{\psi}(t)\rangle\,
\mathrm{d}N(t)\,,
\end{eqnarray}
and insert it into the Taylor expansion of $\mathrm{d}\big(\langle
\widehat{\psi}(t)| \widehat{\psi}(t)\rangle^{-1/2}\big)$ which
yields
\begin{eqnarray}\label{non2}
\mathrm{d}\big(\langle\widehat{\psi}(t)|\widehat{\psi}(t)
\rangle^{-1/2}\big)= \rule{55ex}{0ex}\nonumber
\\=\langle\widehat{\psi}(t)|
\widehat{\psi}(t)\rangle^{-1/2}\left\{
\frac{1}{2}\left(\mu\langle b^{\dagger}b\rangle_{t}+2\sqrt{\mu}\,
\mathrm{Re}\big(\langle b\rangle_{t}\overline{f(t)}\big)+|f(t)|^2
\right)\,\mathrm{d}t\right.\nonumber\\
\left.+\left[\left(\mu\langle
b^{\dagger}b\rangle_{t}+2\sqrt{\mu}\,\mathrm{Re}\big(\langle
b\rangle_{t}\overline{f(t)}\big)+|f(t)|^2\right)^{-1/2}-1\right]
\mathrm{d}N(t)\right\}\,,
\end{eqnarray}
where $\langle . \rangle_{t}=\langle \widehat{\varphi}(t)|(.)
\widehat{\varphi}(t)\rangle$. By eqs. (\ref{fillinear1}), (\ref{non2})
and the Ito formula
\begin{equation}
\mathrm{d}\widehat{\varphi}(t)\;=\; \mathrm{d}\left(\langle
\widehat{\psi}(t)| \widehat{\psi}(t)\rangle^{-1/2}\right)
\widehat{\psi}(t)+ \langle\widehat{\psi}(t)|
\widehat{\psi}(t)\rangle^{-1/2}\mathrm{d}\widehat{\psi}(t)+
\mathrm{d} \left(\langle\widehat{\psi}(t)|\widehat{\psi}(t)
\rangle^{-1/2}\right) \mathrm{d}\widehat{\psi}(t)
\end{equation}
we obtain the nonlinear filtering equation of the form
\begin{eqnarray}\label{filnon1}
\mathrm{d}\widehat{\varphi}(t)&\;=\;& \bigg(-K-\sqrt{\mu}\,
b^{\dagger}\,f(t)+\mu/2\, \langle b^{\dagger}b\rangle_{t}
+\sqrt{\mu}\,\mathrm{Re}\big(\langle b
\rangle_{t}\overline{f(t)}\big)\bigg)\widehat{\varphi}(t)
\mathrm{d}t\nonumber \\ && +\left(\frac{\sqrt{\mu}\,b
+f(t)}{\sqrt{\mu \langle b^{\dagger}b\rangle_{t}
+2\sqrt{\mu}\,\mathrm{Re}\big( \langle
b\rangle_{t}\overline{f(t)}\big)+|f(t)|^{2}}}
-I\right)\widehat{\varphi}(t)\,\mathrm{d}N(t)\,,\;\;
\widehat{\varphi}(0)=\psi\,.
\end{eqnarray}

If the initial state of the system $\mathcal{S}$ is given by a density
matrix
$\rho$, then the posterior normalized density matrix $\widehat{\rho}(t)$
satisfies the stochastic equation
\begin{eqnarray}\label{QSDE4}
\mathrm{d}\widehat{\rho}(t)&=&
-\frac{\mathrm{i}}{\hbar}[H,\widehat{\rho}(t)]\,\mathrm{d}t-
\frac{\mu}{2}\{b^{\dagger}b, \widehat{\rho}(t)\}\,\mathrm{d}t +\mu\,
b\,\widehat{\rho}(t)\,b^{\dagger}\,\mathrm{d}t+\sqrt{\mu}\,[b
\overline{f(t)}-
b^{\dagger}f(t),\widehat{\rho}(t)]\,\mathrm{d}t\nonumber\\
&&+\left(\frac{\big(\sqrt{\mu}\,b+f(t)\big)\widehat{\rho}(t)
\big(\sqrt{\mu}\,b^{\dagger}+ \overline{f(t)}\big)}{\mu
\langle b^{\dagger}b\rangle_{t}+2\sqrt{\mu}\,\mathrm{Re} \big(\langle b
\rangle_{t}\overline{f(t)}\big)+|f(t)|^{2}} -\widehat{\rho}(t)\right)
\nonumber\\
&&\;\;\;\;\times \bigg(\mathrm{d}N(t)- \mu\langle
b^{\dagger}b\rangle_{t}\,\mathrm{d}t- 2\sqrt{\mu}\,\mathrm{Re}\,
\big(\langle b\rangle_{t}\overline{f(t)}\big)\,\mathrm{d}t-|f(t)|^2
\mathrm{d}t\bigg)\,,
\end{eqnarray}
where $\{c,d\}=cd+dc$. The validity of the above equation can be checked
by differentiating $\widehat{\rho}(t)=|
\widehat{\varphi}(t)\rangle\langle\widehat{\varphi}(t)|$.

By taking the stochastic mean of (\ref{QSDE4}), we obtain the master
equation
\begin{equation}\label{mas}
\frac{\mathrm{d}}{\mathrm{d}t}{\sigma}(t)\;=\;-\frac{\mathrm{i}}
{\hbar}[H,\sigma(t)] - \frac{\mu}{2}\{b^{\dagger}b, \sigma(t)\}+
\mu\, b\,\sigma(t)b^{\dagger}\,+\sqrt{\mu}\,[b\,\overline{f(t)}
-b^{\dagger}f(t)\,, \sigma(t)]
\end{equation}
for the prior state
\begin{equation}
\sigma(t)\;=\;\langle\widehat{\rho}(t)\rangle_{st}\,,
\end{equation}
since the posterior mean value of $\mathrm{d}N(t)$ is given by
\begin{equation}
\langle\mathrm{d}N(t)\rangle(\tau)\;=\;\mu \langle b^{\dagger}b
\rangle_{t}\,\mathrm{d}t+2\sqrt{\mu}\, \mathrm{Re}\big(
\langle b\rangle_{t}\overline{f(t)}\big)\,\mathrm{d}t+|f(t)|^2\,
\mathrm{d}t\,.
\end{equation}

Let us note that for $f(t)=0$ the equation (\ref{QSDE4}) takes the
form of a quantum filtering equation for the reservoir prepared
initially in the vacuum state, derived, for instance, in
\cite{Bel89}. In the model worked out by Hudson and Parthasarathy,
the range of frequency of the reservoir extends from $-\infty$ to
$+\infty$. A coherent state of the continuum mode distribution,
having a very narrow spectral width, is here only an analogue of a
single-mode laser field and $f(t)=\lambda \exp(-\mathrm{i}\omega
t+\mathrm{i}\phi)\,,\lambda\geq 0$ \cite{CLG87}. In the considered
experiment the laser light constitutes a coherent signal which
stimulates the system $\mathcal{S}$ and reaches the detector. We
observe the interference between the laser light and the light
emitted by the system $\mathcal{S}$ in the same channel. Because
of the presence of the terms corresponding to the stimulation of
the system $\mathcal{S}$, we cannot take the limit
$|f|\rightarrow\infty$ in (\ref{QSDE4}).

\section{Heterodyne measurement. Transition from counting to diffusion
process}

In the heterodyne detection scheme, depicted in Fig. 1, the
measured signal is superposed with an auxiliary laser field (local
oscillator) by means of the beam splitter  \cite{Bar90, Bar06,
CLG87}. For a lossless beam splitter of transmissivity $T$, the
field reaching detector can be written as
\begin{equation}
A_{\mathrm{mix}}(t)\;=\;\sqrt{T}A^{\mathrm{out}}(t)+\mathrm{i}\,
\sqrt{1-T}\,A_{\mathrm{lo}}(t)\,.
\end{equation}
The operators $A_{\mathrm{lo}}(t)$, $A_{\mathrm{lo}}^{\dagger}(t)$
represent the local oscillator and satisfy the commutation relations
(\ref{canrel}). The auxiliary field does not interact with $\mathcal{S}$
and we assume that its initial state is given by the coherent vector
$\iota(f_{\mathrm{lo}})$.

The output generating operator for the ordinary heterodyne measurement
is defined as
\begin{equation}
\mathrm{G}^{\mathrm{out}}(k,t)\;=\;\langle \iota(f_{\mathrm{lo}})|
\exp\bigg\{\int\limits_{0}^{t}\ln k(t^{\prime})\,
\mathrm{d}{A_{\mathrm{mix}}}(t^{\prime})\dot{A}_{\mathrm{mix}}
(t^{\prime})\bigg\} \iota(f_{\mathrm{lo}})\rangle\,,
\end{equation}
where $\dot{A}_{\mathrm{mix}}(t)=\mathrm{d}A_{\mathrm{mix}}(t)/
\mathrm{d}t$. In this way the degrees of freedom of the auxiliary field
have been eliminated from our description. One can check using
(\ref{diff}) and (\ref{Itotable}) that $\big(\mathrm{d}
{A_{\mathrm{mix}}}(t)\dot{A}_{\mathrm{mix}}(t)\big)^{2}=
\mathrm{d} {A_{\mathrm{mix}}}(t)\dot{A}_{\mathrm{mix}}(t)$. The
assumptions that the field $A^{\mathrm{out}}(t)$ is not lost, i.e.
$T\rightarrow 1$, together with $|f_{\mathrm{lo}}|\rightarrow
\infty$, such that the product
$(1-T)|f_{\mathrm{lo}}|^{2}:=\varepsilon^{-2}$ is fixed
\cite{GarCol85, Car93}, lead to the formula
\begin{equation}\label{genoper}
\mathrm{G}^{\mathrm{out}}(k,t)\;=\;\exp\bigg\{\int\limits_{0}^{t}
\ln k(t^{\prime})\,\mathrm{d}{{Y}}^{\mathrm{out}}(t^{\prime})\bigg\}\,
\end{equation}
with the output counting process ${Y}^{\mathrm{out}}(t)$ of the form
\begin{equation}\label{coupro}
{Y}^{\mathrm{out}}(t)\;=\;\int\limits_{0}^{t}\mathrm{d}
\mathit{\Lambda}^{\mathrm{out}}(t^{\prime})+
\frac{r(t^{\prime})}{\varepsilon}\,
\mathrm{d}A^{\mathrm{out}\dagger} (t^{\prime})+
\frac{\overline{r(t^{\prime})}}{\varepsilon}\,
\mathrm{d} A^{\mathrm{out}}(t^{\prime})+
\frac{1}{\varepsilon^{2}}\,\mathrm{d}t^{\prime}\,,
\end{equation}
$r(t)$ is a complex function with modulus $|r(t)|=1$.

Let us note that from the relations (\ref{canrel}) it follows that
${Y}(t) = U(t){Y}^{\mathrm{out}}(t)U^{\dagger}(t)$ commutes with
${Y}(t^{\prime})$ for any $t$ and $t^{\prime}$. Hence, taking into
account that \cite{Bar86}
$U(t){Y}^{\mathrm{out}}(t^{\prime})={Y}(t^{\prime})U(t)$ one gets
\begin{equation}\label{nondem}
[{Y}^{\mathrm{out}}(t),{Y}^{\mathrm{out}}(t^{\prime})]\;=\;
0\;\;\;\;\;\; \forall\, t,\,t^{\prime}\geq 0\,.
\end{equation}
The property (\ref{nondem}) allows us to treat the process
(\ref{coupro}) as a classical one.

The generating map $\mathrm{g}(k,t)$ for the operator
$\mathrm{G}^{\mathrm{out}}(k,t)$ defined by (\ref{genoper}) satisfies
the differential equation
\begin{eqnarray}\label{genmap2}
\frac{\mathrm{d}}{\mathrm{d}t}\mathrm{g}(k,t)\big[Z\big] \;=\;
\mathrm{g}(k,t)\left[ -\left(K+ \sqrt{\mu}\,b^{\dagger}f(t)+
\sqrt{\mu}\,b\,\overline{r(t)}/\varepsilon+
1/2\,|f(t)+r(t)/\varepsilon|^{2}\right)^{\dagger}Z\right.\nonumber\\
-Z\left(K+ \sqrt{\mu}\,b^{\dagger}f(t)+\sqrt{\mu}\,b\,\overline{r(t)}/
\varepsilon+
1/2\,|f(t)+r(t)/\varepsilon|^{2}\right)\\
\left.+k(t)\left(\sqrt{\mu}\,b^{\dagger}+\overline{f(t)} +
\overline{r(t)}/\varepsilon \right)Z\left(\sqrt{\mu}\,b+
f(t)+r(t)/\varepsilon\right)\right]\,\nonumber
\end{eqnarray}
with $g(k,0)[Z]=Z$. The solution to this equation can be written in the
form
\begin{equation}
\mathrm{g}(k,t)[Z] \;=\; \int\limits_{\kappa\in \Omega^{t}}k(\kappa)
V^{\dagger}(\kappa\,|\,t)\,ZV(\kappa\,|\,t)\mathrm{d}\kappa\,,
\end{equation}
where $\kappa$ denotes the counting trajectory of registered photons
for heterodyne measurement, $\Omega^{t}=\bigcup\limits_{n=0}^{\infty}
\{\kappa \subset[0,t): |\kappa|=n\}$, and
\begin{equation}
V(\kappa\,|\,t)\;=\;\mathrm{e}^{-R(t)}S(t_{n})\,... \,S(t_{1})\,,
\end{equation}
where 
\begin{equation}
S(t)\;=\;\mathrm{e}^{R(t)}\left( \sqrt{\mu}\, b+f(t)+
\frac{r(t)}{\varepsilon}\right) \mathrm{e}^{-R(t)}\,,
\end{equation}
\begin{equation}
R(t)\;= \;Kt+\int\limits_{0}^{t}\,\bigg(\sqrt{\mu}\,
b^{\dagger}f(t^{\prime})+ \sqrt{\mu}\,
b\,\overline{r(t^{\prime})}/\varepsilon+1/2\,|f(t^{\prime})+
r(t^{\prime})/\varepsilon|^{2} \bigg) \mathrm{d}t^{\prime}\,.
\end{equation}
Here, the stochastic propagator $\widehat{V}(t)(\kappa)=V(\kappa\,|\,t)$
given by the formula
\begin{eqnarray}
\widehat{V}(t)\;=\;\mathrm{e}^{-R(t)}
\sum_{n=0}^{\infty}\,\int\limits_{0}^{t}\mathrm{d}t_{n}
\int\limits_{0}^{t_{n}}\mathrm{d}t_{n-1} \ldots
\int\limits_{0}^{t_{2}}\mathrm{d}t_{1}\, S^{\prime}(t_{n})\,\ldots \,
S^{\prime}(t_{1})\prod_{i=1}^{n} \mathrm{d}Y(t_{i})\,,
\end{eqnarray}
where
\begin{equation}
S^{\prime}(t)\;=\;\mathrm{e}^{R(t)}\left( \sqrt{\mu}\, b+f(t)+
\frac{r(t)}{\varepsilon}-I\right)\mathrm{e}^{-R(t)}\,,
\end{equation}
and $Y(t)$ is a random variable such that $\mathrm{d}Y(t)(\kappa)=1$ for
$t\in \kappa$ and $\mathrm{d}Y(t)(\kappa)=0$ for $t \notin \kappa$,
satisfies the equation
\begin{equation}
\mathrm{d}\widehat{V}(t)\;=\;-R(t)\widehat{V}(t)\mathrm{d}t+
\bigg(\sqrt{\mu}\,b+f(t)+\frac{r(t)}{\varepsilon}-I\bigg) \widehat{V}(t)
\mathrm{d}Y(t)\,,\;\;\;\;\widehat{V}(0)\;=\;I\,.
\end{equation}
Hence, we get for the posterior unnormalized wave function
$\widehat{\psi}(t)=\widehat{V}(t)\psi$ the linear Belavkin equation of
the form
\begin{equation}\label{filtlin2}
\mathrm{d}\widehat{\psi}(t)\;=\;-R(t)\widehat{\psi}(t)\mathrm{d}t+
\bigg(\sqrt{\mu}\,b+f(t)+\frac{r(t)}{\varepsilon}-I\bigg)
\widehat{\psi}(t) \mathrm{d}Y(t)\,,\;\;\;\;\widehat{\psi}(0)=\psi\,.
\end{equation}
From the equation (\ref{filtlin2}) one can easily derive the nonlinear
filtering equation for the normalized vector
$\widehat{\varphi}(t)=\langle\widehat{\psi}(t) |\widehat{\psi}(t)
\rangle^{-1/2}\widehat{\psi}(t)$,
\begin{eqnarray}\label{nonlin2}
\mathrm{d}\widehat{\varphi}(t)&\;=\;&-\left(K+\sqrt{\mu}\,
b^{\dagger}f(t)+ \sqrt{\mu}\,b\,\overline{r(t)}/\varepsilon \right)
\widehat{\varphi}(t) \mathrm{d}t\rule{10ex}{0ex}\nonumber\\
&&+\left(\mu/2\, \langle b^{\dagger}b\rangle_{t}+\sqrt{\mu}\,
\mathrm{Re}\,\left(\langle b\rangle_{t} \left(\overline{f(t)}+
\overline{r(t)}/\varepsilon\right)\right)\right)
\widehat{\varphi}(t) \mathrm{d}t\nonumber \\
&&
+\bigg(\left(\mu\langle b^{\dagger}b\rangle_{t}+2\sqrt{\mu}\,
\mathrm{Re}\,\left(\langle b\rangle_{t}\left(\overline{f(t)}+
\overline{r(t)}/\varepsilon\right)\right)
+\left|f(t)+r(t)/\varepsilon\right|^2\right)^{-1/2} \bigg.\nonumber\\
&&\rule{5ex}{0ex}\times\bigg.\left(\sqrt{\mu}\,b+f(t)+r(t)/
\varepsilon\right)-I\bigg) \,\widehat{\varphi}(t)\,\mathrm{d}Y(t)\,,
\end{eqnarray}
where $\langle . \rangle_{t}=\langle \widehat{\varphi}(t)|(.)
\widehat{\varphi}(t)\rangle$ is the posterior mean value of an operator
of the system $\mathcal{S}$. Furthermore, the filtering equation for
the posterior density matrix corresponding to (\ref{nonlin2}) has
the form
\begin{eqnarray}\label{nonlinmat2}
\lefteqn{\mathrm{d}\widehat{\rho}(t)=
\bigg(-\frac{\mathrm{i}}{\hbar}[H,\widehat{\rho}(t)]-
\frac{\mu}{2}\{b^{\dagger}b, \widehat{\rho}(t)\}+\mu b\,
\widehat{\rho}(t)b^{\dagger}+\sqrt{\mu}\,[b\overline{f(t)}
-b^{\dagger}f(t),\widehat{\rho}(t)]\bigg)\mathrm{d}t}\nonumber\\
&& +\left(\varepsilon \mu b\,\widehat{\rho}(t)b^{\dagger}+
\sqrt{\mu}\,b\left(\varepsilon
\overline{f(t)}+\overline{r(t)}\right) \widehat{\rho}(t)+
\widehat{\rho}(t)\sqrt{\mu}\,b^{\dagger}\left(\varepsilon
f(t)+r(t)\right)\right.\nonumber\\&&\rule{4ex}{0ex}\left.
-\varepsilon\mu\langle b^{\dagger}b\rangle_{t}\widehat{\rho}(t)-
2\sqrt{\mu}\,\mathrm{Re}\,\left(\langle b\rangle_{t}\left(\varepsilon
\overline{f(t)}+\overline{r(t)}\right) \right)\widehat{\rho}(t)
\right)\nonumber\\&&\times \bigg(\varepsilon\mathrm{d}\,Y(t)-
\varepsilon\mu\langle b^{\dagger}b\rangle_{t}\mathrm{d}t-
2\sqrt{\mu}\,\mathrm{Re}\,\left(\langle b\rangle_{t}\left(\varepsilon
\overline{f(t)}+\overline{r(t)}\right) \right)\mathrm{d}t-
\varepsilon^{-1}\left|\varepsilon f(t)+r(t)\,\right|^2\mathrm{d}t
\bigg)\nonumber\\
&&\times \left(\varepsilon^{2}\mu\langle b^{\dagger}b\rangle_{t}+
2\sqrt{\mu}\,\mathrm{Re}\,\left(\langle b\rangle_{t}\,
\left(\varepsilon^2\overline{f(t)}+\varepsilon\,\overline{r(t)}
\right)\right)+ \left|\varepsilon f(t)+ r(t)\,\right|^2\right)^{-1}\,.
\end{eqnarray}
Of course, if we put $r(t)=0$ in the above equations, we get the
results obtained in the previous section for a direct observation of
the output field. Moreover, since the posterior mean value of
$\mathrm{d}Y(t)$ is given by the formula
\begin{equation}
\langle\mathrm{d}Y(t)\rangle(\kappa)\;=\;\left(\mu\langle
b^{\dagger}b\rangle_{t}+ 2\sqrt{\mu}\,\varepsilon^{-1}\mathrm{Re}\,
\left(\langle b\rangle_{t}\left(\varepsilon\overline{f(t)}+
\overline{r(t)}\right) \right)+\varepsilon^{-2}\left|\varepsilon f(t)+
\overline{r(t)}\,\right|^2\right)\mathrm{d}t\,,
\end{equation}
the averaging of (\ref{nonlinmat2}) on the past leads to the master
equation of the form (\ref{mas}).

Now we can consider the linear transformation of the counting process
$Y(t)$ \cite{BarBel91},
\begin{equation}\label{transform}
\mathrm{d}W^{\varepsilon}(t)\;=\;\varepsilon\,\mathrm{d}Y(t)-
\frac{\mathrm{d}t}{\varepsilon}\,,
\end{equation}
leading in the limit $\varepsilon\rightarrow 0$ to diffusion
observation. One can easily check that
\begin{eqnarray}
\mathrm{d}W^{\varepsilon}(t)\mathrm{d}W^{\varepsilon}(t)\;=\;
\varepsilon^{2}\mathrm{d}Y(t)\;=\;\varepsilon\,
\mathrm{d}W^{\varepsilon}(t)+\mathrm{d}t\,,\;\;\;
\mathrm{d}W^{\varepsilon}(t)\mathrm{d}t\;=\;0\,,
\end{eqnarray}
such that for $W(t)=\lim\limits_{\varepsilon \rightarrow 0}
W^{\varepsilon}(t)$ we obtain the following Ito rules:
\begin{equation}
\mathrm{d}W(t)\mathrm{d}W(t)\;=\;\mathrm{d}t\,,\;\;\;
\mathrm{d}W(t)\mathrm{d}t\;=\;0\,.
\end{equation}
Taking $\varepsilon \rightarrow 0$ one can get from (\ref{nonlinmat2})
the filtering equation for the diffusion observation
\begin{eqnarray}\label{diffu}
\mathrm{d}\widehat{\rho}(t)\;&=&\;\bigg(
-\frac{\mathrm{i}}{\hbar}[H,\widehat{\rho}(t)]-\frac{\mu}{2}
\{b^{\dagger}b,\widehat{\rho}(t)\}+\mu b\widehat{\rho}(t)
b^{\dagger}+\sqrt{\mu}\,[b\overline{f(t)} -b^{\dagger}f(t),
\widehat{\rho}(t)]\bigg)\mathrm{d}t\nonumber \\
&& +\left(\sqrt{\mu}\,\overline{r(t)}\left(b -
\langle b\rangle_{t}\right)\widehat{\rho}(t)+
\sqrt{\mu}\,r(t)\widehat{\rho}(t)\left(b^{\dagger}-
\langle b^{\dagger}\rangle_{t}\right)\right)\nonumber \\ &&
\times\bigg(\mathrm{d}W(t)- 2\,\mathrm{Re}\,\left(\sqrt{\mu}\,
\langle b\rangle_{t}\overline{r(t)}+f(t)\overline{r(t)}
\right)\,\mathrm{d}t\bigg)\,.
\end{eqnarray}

Let us note that eq. (\ref{diffu}) transforms pure states into pure ones.
In such a case, for\break $\widehat{\rho}(t)=|\widehat{\varphi}(t)
\rangle \langle\widehat{\varphi}(t)|$ eq. (\ref{diffu}) takes the form
\begin{eqnarray}
\lefteqn{\mathrm{d}\widehat{\varphi}(t)\;=\; \bigg(-K+\sqrt{\mu}\,b
\overline{f(t)}-\sqrt{\mu}\,b^{\dagger}f(t)+\mu\,\langle b^{\dagger}
\rangle_{t}b - \frac{\mu}{2}\,|\langle b\rangle_{t}|^{2} \bigg)
\widehat{\varphi}(t)\,\mathrm{d}t}\nonumber\\ &&
+\sqrt{\mu}\,\overline{r(t)}\,\big(b-\langle b\rangle_{t}\big)\,
\widehat{\varphi}(t)\bigg(\mathrm{d}W(t)- 2\,\mathrm{Re}\,\left(
\sqrt{\mu}\,\langle b\rangle_{t}\overline{r(t)}+f(t)\overline{r(t)}
\right)\,\mathrm{d}t\bigg)\,.
\end{eqnarray}

Let us note that if we take $f(t)=0$, then all the obtained
formulae reduce to the known results for the reservoir initially
prepared in the vacuum state \cite{BarBel91}. Finally it is worth
to mention that modelling of the evolution of a quantum state
conditioned by the results of a continuous measurement is the
first step towards quantum feedback control \cite{Bel99, WisMil00}
with a coherent source used as a control field \cite{BelEdw08}. As
the choice of the system's operators does not interfere with the
derivation of the filtering equations for a measurement of a
coherent channel, the results of the paper can be easily
generalized -- in the obtained filtering equations, the system's
hamiltonian and  system's coupling operator can be replaced with
arbitrary ones.

\newpage 

\noindent {\bf Figure captions}

\noindent Fig. 1. The heterodyne setup

\newpage

\begin{center}
\begin{figure}[t]\label{Fig1}
\begin{picture}(220,100)(50,1)
\put(80,60){\vector(1,0){50}}
\put(130,60){\line(1,0){20}}
\put(85,65){\small{input field}}
\put(150,50){\framebox(70,25){\small{system $\mathcal{S}$}}}
\put(220,60){\vector(1,0){50}}
\put(270,60){\line(1,0){20}}
\put(225,65){\small{output field}}
{\thicklines
\put(270,40){\line(1,1){40}}}
\put(287,60){\vector(1,0){50}}
\put(337,60){\line(1,0){20}}
\put(357,50){\framebox(70,25){\small{photodetector}}}
\put(290,60){\vector(0,1){40}}
\put(290,100){\line(0,1){10}}
\put(312,82){\shortstack{\small{beam}\\ \small{splitter}}}
\put(290,0){\vector(0,1){40}}
\put(290,40){\line(0,1){20}}
\put(292,20){\small{local oscillator}}
\end{picture}
\noindent\caption{\small The heterodyne setup}
\end{figure}
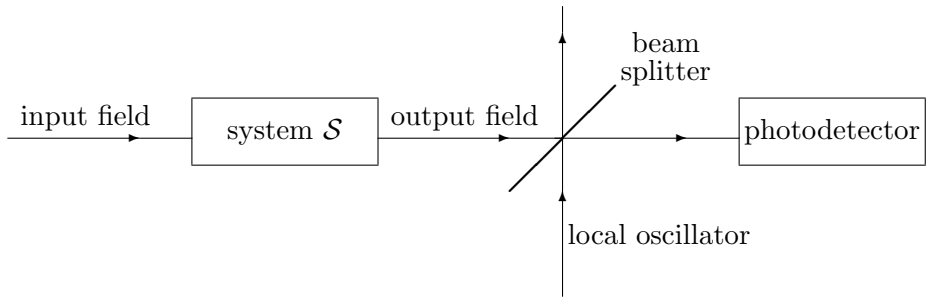
\end{center}
\end{document}